\documentclass[a4paper,aps,twocolumn,pre,showpacs,superscriptaddress,floatfix,nofootinbib]{revtex4-1}
\usepackage{graphicx,amsmath}
\usepackage{wasysym,pifont}
\usepackage{epsfig}
\usepackage{newlfont,bm,dcolumn,longtable}
\usepackage{amssymb,amsfonts,amsmath}
\usepackage{color}

\begin{document}

\newcommand{\be}{\begin{equation}}
\newcommand{\ee}{\end{equation}}
\newcommand{\bea}{\begin{eqnarray}}
\newcommand{\eea}{\end{eqnarray}}
\newcommand{\<}{\langle}
\renewcommand{\>}{\rangle}
\newcommand{\qea}{q_{\scriptscriptstyle{\rm EA}}}
\newcommand{\teff}{t_{\rm eff}}
\newcommand{\thav}[1]{\langle #1 \rangle}
\newcommand{\smav}[1]{\overline{#1}}
\newcommand{\average}[1]{\smav{\thav{#1}}}
\title{
Sample-to-sample fluctuations of the overlap distributions in the three-dimensional Edwards-Anderson spin glass
}

\author{R.~A.~Ba\~nos}
\affiliation{Instituto de Biocomputaci\'on y  F\'{\i}sica de
Sistemas Complejos (BIFI),\\
Facultad de Ciencias, Universidad de Zaragoza, 50009 Zaragoza, Spain}

\author{A.~Cruz}
\affiliation{Departamento de F\'{\i}sica Te\'orica,
Facultad de Ciencias, \\
Universidad de Zaragoza, 50009 Zaragoza, Spain}
\affiliation{Instituto de Biocomputaci\'on y  F\'{\i}sica de
Sistemas Complejos (BIFI),\\
Facultad de Ciencias, Universidad de Zaragoza, 50009 Zaragoza, Spain}

\author{L.A.~Fernandez}
\affiliation{Departamento de F\'{\i}sica Te\'orica, Facultad de
  Ciencias F\'{\i}sicas,\\
Universidad Complutense de Madrid, 28040 Madrid, Spain}
\affiliation{Instituto de Biocomputaci\'on y  F\'{\i}sica de
Sistemas Complejos (BIFI),\\
Facultad de Ciencias, Universidad de Zaragoza, 50009 Zaragoza, Spain}

\author{J.M.~Gil-Narvion}
\affiliation{Instituto de Biocomputaci\'on y  F\'{\i}sica de
Sistemas Complejos (BIFI),\\
Facultad de Ciencias, Universidad de Zaragoza, 50009 Zaragoza, Spain}

\author{A.~Gordillo-Guerrero}
\affiliation{Departamento de F\'{\i}sica, Facultad de Ciencias,\\
Universidad de Extremadura, 06071 Badajoz, Spain}
\affiliation{Instituto de Biocomputaci\'on y  F\'{\i}sica de
Sistemas Complejos (BIFI),\\
Facultad de Ciencias, Universidad de Zaragoza, 50009 Zaragoza, Spain}

\author{M.~Guidetti}
\affiliation{Instituto de Biocomputaci\'on y  F\'{\i}sica de
Sistemas Complejos (BIFI),\\
Facultad de Ciencias, Universidad de Zaragoza, 50009 Zaragoza, Spain}

\author{D.~I\~niguez}
\affiliation{Instituto de Biocomputaci\'on y  F\'{\i}sica de
Sistemas Complejos (BIFI),\\
Facultad de Ciencias, Universidad de Zaragoza, 50009 Zaragoza, Spain}
\affiliation{Fundaci\'on ARAID, Diputaci\'on General de Arag\'on, Zaragoza, Spain}

\author{A.~Maiorano}
\affiliation{Dipartimento di Fisica, Sapienza Universit\`a di Roma, 00185 Roma, Italy}
\affiliation{Instituto de Biocomputaci\'on y  F\'{\i}sica de
Sistemas Complejos (BIFI),\\
Facultad de Ciencias, Universidad de Zaragoza, 50009 Zaragoza, Spain}

\author{F.~Mantovani}
\affiliation{Deutsches Elektronen-Synchrotron (DESY),D-15738 Zeuthen, Germany}

\author{E.~Marinari}
\affiliation{Dipartimento di Fisica, INFN Sezione di Roma I, IPCF-CNR UOS Roma,
Sapienza Universit\`a di Roma, 00185 Roma, Italy}

\author{V.~Martin-Mayor}
\affiliation{Departamento de F\'{\i}sica Te\'orica, Facultad de
  Ciencias F\'{\i}sicas,\\
Universidad Complutense de Madrid, 28040 Madrid, Spain}
\affiliation{Instituto de Biocomputaci\'on y  F\'{\i}sica de
Sistemas Complejos (BIFI),\\
Facultad de Ciencias, Universidad de Zaragoza, 50009 Zaragoza, Spain}

\author{J.~Monforte-Garcia}
\affiliation{Instituto de Biocomputaci\'on y  F\'{\i}sica de
Sistemas Complejos (BIFI),\\
Facultad de Ciencias, Universidad de Zaragoza, 50009 Zaragoza, Spain}

\author{A.~Mu\~noz-Sudupe}
\affiliation{Departamento de F\'{\i}sica Te\'orica, Facultad de
  Ciencias F\'{\i}sicas,\\
Universidad Complutense de Madrid, 28040 Madrid, Spain}

\author{D.~Navarro}
\affiliation{Departamento de Ingenier\'{\i}a, Electr\'onica y Comunicaciones\\
and Instituto de Investigaci\'on en Ingenier\'{\i}a de Arag\'on (I3A)}
\author{G.~Parisi}
\affiliation{Dipartimento di Fisica, INFN Sezione di Roma I, IPCF-CNR UOS Roma,
Sapienza Universit\`a di Roma, 00185 Roma, Italy}

\author{S.~Perez-Gaviro}
\affiliation{Dipartimento di Fisica, Sapienza Universit\`a di Roma, 00185 Roma, Italy}
\affiliation{Instituto de Biocomputaci\'on y  F\'{\i}sica de
Sistemas Complejos (BIFI),\\
Facultad de Ciencias, Universidad de Zaragoza, 50009 Zaragoza, Spain}

\author{F.~Ricci-Tersenghi}
\affiliation{Dipartimento di Fisica, INFN Sezione di Roma I, IPCF-CNR UOS Roma,
Sapienza Universit\`a di Roma, 00185 Roma, Italy}

\author{J.~J.~Ruiz-Lorenzo}
\affiliation{Departamento de F\'{\i}sica, Facultad de Ciencias,\\
Universidad de Extremadura, 06071 Badajoz, Spain}
\affiliation{Instituto de Biocomputaci\'on y  F\'{\i}sica de
Sistemas Complejos (BIFI),\\
Facultad de Ciencias, Universidad de Zaragoza, 50009 Zaragoza, Spain}

\author{S.F.~Schifano}
\affiliation{Dipartimento di Matematica, Universit\`a di Ferrara and INFN, Ferrara, Italy}

\author{B.~Seoane}
\affiliation{Departamento de F\'{\i}sica Te\'orica, Facultad de
  Ciencias F\'{\i}sicas,\\
Universidad Complutense de Madrid, 28040 Madrid, Spain}
\affiliation{Instituto de Biocomputaci\'on y  F\'{\i}sica de
Sistemas Complejos (BIFI),\\
Facultad de Ciencias, Universidad de Zaragoza, 50009 Zaragoza, Spain}

\author{A.~Taranc\'on}
\affiliation{Departamento de F\'{\i}sica Te\'orica,
Facultad de Ciencias, \\
Universidad de Zaragoza, 50009 Zaragoza, Spain}
\affiliation{Instituto de Biocomputaci\'on y  F\'{\i}sica de
Sistemas Complejos (BIFI),\\
Facultad de Ciencias, Universidad de Zaragoza, 50009 Zaragoza, Spain}

\author{R.~Tripiccione}
\affiliation{Dipartimento di Fisica, Universit\`a di Ferrara and INFN, Ferrara,
Italy)}

\author{D.~Yllanes}
\affiliation{Departamento de F\'{\i}sica Te\'orica, Facultad de
  Ciencias F\'{\i}sicas,\\
Universidad Complutense de Madrid, 28040 Madrid, Spain}
\affiliation{Instituto de Biocomputaci\'on y  F\'{\i}sica de
Sistemas Complejos (BIFI),\\
Facultad de Ciencias, Universidad de Zaragoza, 50009 Zaragoza, Spain}
\date{\today}

\date{\today}

\begin{abstract}
We study the sample-to-sample fluctuations of the overlap probability densities 
from large-scale equilibrium simulations of the three-dimensional Edwards-Anderson 
spin glass below the critical temperature. Ultrametricity, Stochastic Stability and Overlap Equivalence
impose constraints on the moments of the overlap probability densities
that can be tested against numerical data.
We found small deviations from the Ghirlanda-Guerra predictions, which get smaller as system size 
increases.
We also focus on the shape of the overlap distribution, comparing the numerical data to a mean-field-like prediction
in which finite-size effects are taken into account by substituting delta functions with broad peaks.
\end{abstract}


\maketitle

\section{INTRODUCTION}
\label{sec:intro}

Spin glasses are model glassy systems which have been studied for
decades and have become a paradigm for a broad class of scientific applications.
They not only provide a mathematical model for disordered alloys and
their striking low-temperature properties (slow dynamics,
age-dependent response), but they have also been the test-ground for
new ideas in the study of other complex systems, such as structural
glasses, colloids, econophysics, and combinatorial optimization
models. The non-trivial phase-space structure of the mean-field
solution to spin glasses~\cite{SK,Parisi80,MPVbook} encodes many
properties of glassy behavior.

Whether the predictions of the mean-field solutions correctly describe
the properties of finite-range spin-glass models (and of their
experimental counterpart materials) is a long-debated question.  The
Droplet Model describes the spin glass phase in terms of a unique
state (apart from a global inversion symmetry) and predicts a
(super-universal) coarsening dynamics for the off-equilibrium regime.~\cite{DM} Moreover, there is no spin glass
transition in presence of any external magnetic field.  On the other side, the
Replica Symmetry Breaking scenario~\cite{MPVbook,MPRRZ}, based on the mean
field prediction, describes a complex free-energy landscape and a non-trivial
order parameter distribution in the thermodynamic limit; the dynamics is
critical at all temperatures in the spin-glass phase.  The spin glass
transition temperature is finite also in presence of small magnetic fields;
the search for the de Almeida-Thouless line $T_\text{c}(h)$ is the purpose of
many numerical experiments (see, for example, Ref.~\onlinecite{SYAT}).

From the theoretical perspective, the last decade has seen a strong
advance in the understanding of the properties of the mean-field
solution: its correctness has been rigorously proved thanks to the
introduction of new concepts and tools, like stochastic stability or
replica and overlap equivalence
\cite{GuerraSS,AizenmanContucci,GhirlandaGuerra,Parisi98,Talagrand}.
Besides, numerical simulation has been the methodology of choice when
investigating finite-range spin glasses, even if the computational
approach is severely plagued by the intrinsic properties (slow
convergence to equilibrium, slowly growing correlation lengths) of the
simulated system's (thermo)dynamics. In this respect, a
Moore-law-sustained improvement in performance of devices for
numerical computation and new emerging technologies in the last years
has allowed for very fast-running implementation of standard simulation
techniques. By means of the non-conventional computer
Janus~\cite{JANUS_SH} we have been able to collect high-quality
statistics of equilibrium configurations of three-dimensional
Edwards-Anderson spin glasses, well beyond what would
have been possible on conventional PC clusters.

Theoretical predictions and Janus numerical data have been compared in
detail in Refs.~\onlinecite{janusPRL} and~\onlinecite{EAPTJANUS}. One
of the main results presented therein is that equilibrium properties at
a given finite length scale correspond to out-of-equilibrium
properties at a given finite time scale. On experimentally accessible
scales (order $10^4$ seconds waiting times corresponding to order
$10^2$ lattice sizes) the Replica Symmetry Breaking picture turns out
as the only relevant effective theory.  Theories in which some of the
fundamental ingredients of the mean-field solutions are lacking
(overlap equivalence in the TNT model~\cite{TNT}, ultrametricity in
the Droplet Model) show inconsistencies when their predictions are
compared to the observed behavior.

In this work we reconsider the analysis of the huge amount of data at our
disposal, focusing on the sample-to-sample fluctuations of 
the distribution of the overlap order parameter. The assumptions of
the mean-field theory allow us to make predictions on the joint
probabilities of overlaps among many real replicas which can be tested
against numerical data for the three-dimensional Edwards-Anderson
model. The structure of the paper is as follows: in section~\ref{sec:simu} we
give some details on the considered spin-glass model
and the performed Monte Carlo simulations. In the
subsequent section we first recall some fundamental concepts such as
stochastic stability, ultrametricity, replica and overlap equivalence
and some predictions on the joint overlap probability densities, and
then present a detailed comparison with numerical data. In
section~\ref{sec:pq} we show how finite-size numerical overlap
distributions compare to the mean-field prediction in which
finite-size effects are appropriately introduced.  We finally present
our conclusions in the last section.

\section{MONTE CARLO SIMULATIONS}
\label{sec:simu}

\subsection{The Model}
\label{subsec:model}
We consider the Edwards-Anderson model~\cite{EAMODEL} in three
dimensions, with Ising spin variables $\sigma_{i}=\pm 1$ and bi\-nary
random quenched couplings $J_{ij}=\pm 1$. Each spin, set on the nodes
of a cubic lattice of size $V=L^3$ ($L$ being the lattice size),
interacts only with its nearest neighbors under periodic boundary
conditions. The Hamiltonian is:
\be
H = - \sum_{\langle i,j \rangle} J_{ij} \sigma_{i} \sigma_{j}\ \ ,
\label{eq:EA}
\ee
where the sum extends over nearest-neighbor lattice sites.
In what follows we are dealing mainly with measures of the \emph{spin
  overlap}
\be
q_{ab}=\frac{1}{L^3}\sum_i\sigma_i^a\sigma_i^b\;,
\label{eq:q}
\ee
where $a$ and $b$ are replica indices, and the sample-dependent
frequencies $N_J(q_{ab})$ with which we estimate the overlap
probability distribution $P_J(q)$ of each sample (we indicate
one-sample quantities by the subscript $J$):
\be
P_J(q_{ab})=\left<
  \delta \left(q_{ab}-\frac{1}{L^3}\sum_i\sigma_i^a\sigma_i^b \right)
\right> \;,
\label{eq:Pjq}
\ee
where $\langle\left(\cdot\cdot\cdot\right)\rangle$ is a thermal average.
In what follows $\overline{\left(\cdot\cdot\cdot\right)}$ denotes average over disorder.

\subsection{Numerical Simulations}
\label{subsec:simu}

\begin{table}[t]
\begin{center}
\begin{tabular}{|c|c|c|c|c|}
\hline
$L$ & $T_\text{min}$ & $T_\text{max}$ & $N_{T}$ & $N_{S}$\\
\hline
   8  & 0.150    & 1.575     & 10     & 4000\\
  16  & 0.479    & 1.575     & 16     & 4000\\
  24  & 0.625    & 1.600     & 28     & 4000\\
  32  & 0.703    & 1.549     & 34     & 1000\\
\hline
\end{tabular}
\end{center}
\caption{A summary of parameters of the simulations we have used in
  this work. For each lattice size, $L$, we considered $N_{S}$
  samples, with four independent real replicas per sample. For the
  Parallel Tempering algorithm, $N_{T}$ temperatures were used between
  $T_\text{min}$ and $T_\text{max}$, uniformly distributed in that range (except
  in the case of $L=8$, in which we have $7$ temperatures uniformly
  distributed between $0.435$ and $1.575$ plus the $3$ temperatures
  $0.150$, $0.245$ and $0.340$). Our MCS consisted of $10$ Heat-Bath
  sweeps followed by $1$ Parallel Tempering update. More detailed
  information regarding these simulations can be found in
  Ref.~\onlinecite{EAPTJANUS}.}
\label{tab:simdetails}
\end{table}

We present an analysis of overlap probability distributions computed on
equilibrium configurations of the three-dimensional Edwards-Anderson
model defined in Eq.~(\ref{eq:EA}). We computed the configurations by means of an intensive Monte Carlo
simulation on the \emph{Janus} supercomputer. Full details of these simulation
can be found in Ref.~\onlinecite{EAPTJANUS}. For easy reference, we summarize
the parameters of our simulations in Table~\ref{tab:simdetails}.
In order to reach such low temperature values, it has been crucial to tailor
the simulation time, on a sample-by-sample basis, through a careful study of the
temperature random-walk dynamics along the parallel tempering simulation.

\section{REPLICA EQUIVALENCE AND ULTRAMETRICITY}
\label{sec:re_and_um}

The Sherrington-Kirkpatrick (SK) model~\cite{SK} is the mean-field
counterpart of model~(\ref{eq:EA}). It is defined by the Hamiltonian
\be
H=\sum_{i\neq j}J_{ij}\sigma_i \sigma_j\;,
\label{eq:SK}
\ee
where the sum now extends to all pairs of $N$ Ising spins and the
couplings $J_{ij}$ are independent and identically-distributed
random variables extracted from a
Gaussian or a bimodal distribution with variance $1/N$. The quenched
average of the thermodynamic potential may be performed by rewriting
the $n$-replicated partition function in terms of an $n\times n$
overlap matrix $Q_{a,b}$ for which the saddle-point approximation
gives the self-consistency equation
\be
Q_{ab}=\langle \sigma^a \sigma^b \rangle\;,
\ee
where the average $\langle(\cdot\cdot\cdot)\rangle$ involves an effective
single-site Hamiltonian in which $Q_{a,b}$ couples the replicas.  The
thermodynamics of model~(\ref{eq:SK}) is recovered in the limit
$n\rightarrow 0$, after averaging over all possible permutations of
replicas.

The overlap probability distribution $P(q)$ is defined in terms of
such an averaging procedure: for any function of the overlap $f(q)$, one
has that
\be
\int dq_{a,b} P(q_{a,b}) f(q_{a,b}) = \lim_{n\rightarrow 0}
\frac{1}{n!}\sum_{p}f(Q_{p(a),p(b)})\;,
\ee
the sum being over permutations $p$ of the $n$ replica indices.  The
assumption of the replica approach is that $P(q)$ defined in this way
is the same as the large-volume limit of the disorder average
$\overline{P_J(q)}$ of the probability distribution of the overlap
defined in Eqs.~(\ref{eq:q}) and (\ref{eq:Pjq}).

The hierarchical solution~\cite{MPVbook} for $Q_{ab}$ is based on
two main assumptions: stochastic stability and ultrametricity. In what
follows we are interested in the consequences of such assumptions when
dealing with a generic random spin system defined by a Hamiltonian
$H_{J}(\sigma)$, where the subscript $J$ summarizes the dependence on
a set of random quenched parameters, e.g., the random couplings in
models~(\ref{eq:EA}) and (\ref{eq:SK}).

\emph{Stochastic stability}~\cite{GuerraSS,AizenmanContucci} in the
replica formalism is equivalent to replica equivalence~\cite{GhirlandaGuerra,Parisi98}:
one-replica observables retain
symmetry under replica permutation even when the replica symmetry is
broken. This property implies that the $n\times n$ overlap matrix for
an $n$-replicated system, satisfies
\be
\label{eq:RE}
0 \equiv \sum_{c}\left[f(Q_{ac})-f(Q_{bc})\right]
\ee
for any function $f$ and any indices $a,b$.  In the framework of the
solution to the mean-field model, this is necessary for having a well
defined free energy~\cite{Parisi80,Parisi98} in the limit
$n\rightarrow 0$.  A consequence of (\ref{eq:RE}) is, given a set of
$n$ real replicas, the possibility of expressing joint probabilities
of $m$ among the $n(n-1)/2$ overlap variables to joint probabilities
for overlaps among a set of up to $m$ replicas.~\cite{Parisi98} The
following relations hold, for instance, in the cases $n=4,m=2$ and
$n=6,m=3$:
\bea
3P\left(q_{12},q_{34} \right) & = & 2P\left(q_{12}\right)
P\left(q_{34}\right) \nonumber \\
\label{eq:PqSS1} 
& + & \delta \left(q_{12}-q_{34}\right) P\left(q_{12}\right)\;,\\
15P\left(q_{12},q_{34},q_{56}\right) & = &
2P\left(q_{12},q_{23},q_{31}\right) \nonumber \\
 & + & 5P\left(q\right)P\left(q^\prime\right)P\left(q^{\prime\prime}\right) \nonumber \\
 & + & 2\delta\left(q-q^\prime\right)P\left(q^\prime\right)P\left(q^{\prime\prime}\right) \nonumber \\
 & + & 2\delta\left(q^\prime-q^{\prime\prime}\right)P\left(q\right)P\left(q^{\prime}\right) \nonumber \\
 & + & 2\delta\left(q-q^{\prime\prime}\right)P\left(q\right)P\left(q^{\prime}\right) \nonumber \\
\label{eq:PqSS2}
 & + &
2\delta\left(q-q^\prime\right)\delta\left(q^\prime-q^{\prime\prime}\right)P\left(q\right)\;,
\eea
where $q\equiv q_{12}$,  $q^\prime\equiv q_{34}$,  $q^{\prime\prime}\equiv
q_{56}$.

Note that  relation~(\ref{eq:PqSS1}) quantifies the fluctuations of the
overlap distribution: even in the limit of very large
volumes, for the joint probability of two independent overlaps,
\be
P(q_{12},q_{34}^\prime) \equiv \overline{ P_J(q_{12},q_{34}^\prime) } \neq
\overline{ P_J(q_{12}) }\;\overline{ P_J(q_{34}^\prime) }\;.
\ee

\emph{Ultrametricity} is the other remarkable feature of the mean-field
solution, stating that when picking up three equilibrium
configurations, either their mutual overlaps are all equal or two are
equal and smaller than the third. A distance can be defined in terms of
the overlap so that all triangles among states are either equilateral
or isosceles.  In terms of overlaps probabilities, the property reads:
\bea
\label{eq:ultratriple}
P(q_{12},q_{23},q_{31}) & = &  \delta(q_{12}-q_{23})\delta(q_{23}-q_{31})B(q_{12}) \\ 
& + & \left[\Theta(q_{12}-q_{23})A(q_{12},q_{23})\delta(q_{23}-q_{31})
  \right. \nonumber\\
& + &\left.\mbox{two perm.}\right]\nonumber
\eea
where $\Theta(x)$ is the Heaviside step function.
By replica equivalence, $A$ and $B$ can be expressed in terms of $P(q)$:~\cite{IPR}
\bea
\label{eq:A}
A(q,q^\prime) & = & P(q)P(q^\prime)\ \ ,\\
\label{eq:B}
B(q) & = & x(q)P(q)\ \ ,\\
\label{eq:x}
x(q) & = & \int_{-q}^qP(q^\prime)dq^\prime\ \ .
\eea
Ultrametricity implies that the joint
probability of overlaps among $n$ replicas, which in principle
depends on $n(n-1)/2$ variables, is a function of only $n-1$
variables. Thus, using replica equivalence, it is reduced to a
combination of joint probabilities of a smaller set of replicas. Note
that $P(q_{12},q_{23},q_{31})$ is the only \emph{non-single-overlap}
quantity appearing in the r.h.s. of Eq.~(\ref{eq:PqSS2}): by combining
replica equivalence and ultrametricity, three-overlap  probabilities
reduce to combinations of single-overlap probabilities.

Stochastic stability, or equivalently replica equivalence, is a quite
general property that should apply also to short-range models, in the
hypothesis that the model is not unstable upon small random long-range
perturbations~\cite{GuerraSS}. Whether short-range models would
feature ultrametricity is a long-debated question, for which direct
inspection by numerical means is the methodology of choice.  It has
been shown~\cite{PRJPA} that, in the hypothesis that the overlap
distribution is non-trivial and fluctuating in the thermodynamic
limit, then ultrametricity is equivalent to the simpler assumption of
\emph{overlap equivalence}, in the sense that it is the unique
possibility when both replica and overlap equivalence hold. Overlap
equivalence states that, in the presence of replica symmetry breaking,
given any local function $A_i(\sigma)$, the generalized overlap
$q_A=N^{-1}\sum_iA_i(\sigma^a)A_i(\sigma^b)$, with $a,b$ indices of
real replicas, does not fluctuate when considering configurations at
fixed spin-overlap~\cite{Athanasiu}: all definitions of the overlap are equivalent.
Assuming that stochastic stability is a very generic property, there
may be violation of ultrametricity only in a situation in which also
overlap equivalence is violated. In this respect, evidence  of overlap
equivalence has been found in both equilibrium and off-equilibrium
numerical simulations of the Edwards-Anderson
model~\cite{EAPTJANUS,ContucciPRL,janusPRL}.

The aim of this work is a numerical study of the sample-to-sample
fluctuations of the overlap distribution; we focus on the sample
statistics of the cumulative overlap probability functions defined by
\be
\label{eq:XJq}
X_J(q) \equiv \int_{-q}^{q}P_J\left(q^\prime \right) dq^\prime \;.
\ee
This is a random variable, since it depends on the random disorder,
and we denote by $\Pi_{q}(X_J)$ its probability distribution.  We
estimate the moments of the $\Pi_q$ distribution as
\bea
X_k(q) &=& \int x^k \Pi_q(x)dx = \overline{\left[ X_{J}\left( q
    \right)\right]^k} \nonumber \\
&=&\overline{\left[ \int_{-q}^{q}P_J\left(q^\prime \right) dq^\prime
  \right]^k}\;,
\label{eq:kthmoms}
\eea
where $P_J\left( q \right)$ are the Monte Carlo overlap frequencies for
a given sample.

Given a set of three independent spin configurations we obtain also the
probability for the three overlaps to be smaller than $q$:
\be
\label{eq:XT}
X_\text{T}(q)=\overline{\int_{-q}^{q}P_J(q_{12},q_{23},q_{31})dq_{12}dq_{23}dq_{31}}
\ee
In the replica equivalence assumption $X_k(q)$ can be expressed in
terms of $X_\text{T}(q)$ and $X_1(q)$; integrating the Ghirlanda-Guerra
relations (\ref{eq:PqSS1},\ref{eq:PqSS2}) up to $k=3$ we have:
\bea
\label{eq:X2SS}
X_2(q) & = &\frac{1}{3}X_1(q)+\frac{2}{3}X_1^2(q) \ \ ,\\
\label{eq:X3SS}
X_3(q) & = &\frac{1}{15}\left[2X_\text{T}(q)+2X_1(q) \nonumber \right.\\ 
         & + & \left. 6X_1^2(q)+5X_1^3(q)\right]\ \ .
\eea
Ultrametricity imposes a further constraint: from relations
(\ref{eq:ultratriple} - \ref{eq:x}) it follows
\be
\label{eq:XTultra}
X_\text{T}(q)=\left[x(q)\right]^2 \equiv X_1^2(q)\ \ ,
\ee
And the quantities (\ref{eq:kthmoms}) become  polynomials in $X_1$
only.  The above relation simply states that, if ultrametricity holds,
the probability of finding three overlaps smaller than $q$ factorizes
to the probability of finding two overlaps independently smaller than
$q$, with the third bound to be equal to at least one of the previous
two.

\begin{figure}[tb]
\begin{center}
\includegraphics[width=0.8\columnwidth]{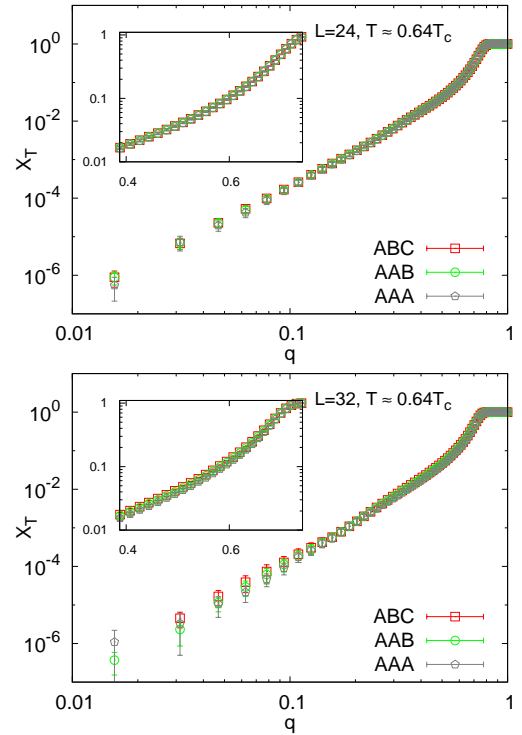}\\
\end{center}
\caption{(Color
  online) The quantity $X_\text{T}$ as defined in the text, as a function of
  $q$ for lattice size $L=24$ (top) and $L=32$ (bottom) at
  temperature $T \simeq 0.64 T_\text{c}$. Insets show a magnified view of the
  region $q \sim 0.6$ (log-log plot). Plots show data for $X_\text{T}$
  computed only with triplets of independent configurations (ABC),
  with triplets in which two configurations belong to the same Monte
  Carlo history (AAB), and triplets in which all configurations come
  from the same Monte Carlo history (AAA). No significant difference
  shows up as long as we take enough uncorrelated configurations from
  the same replica.}
\label{fig:XT}
\end{figure}

\begin{figure}[tb]
\begin{center}
\includegraphics[width=0.8\columnwidth]{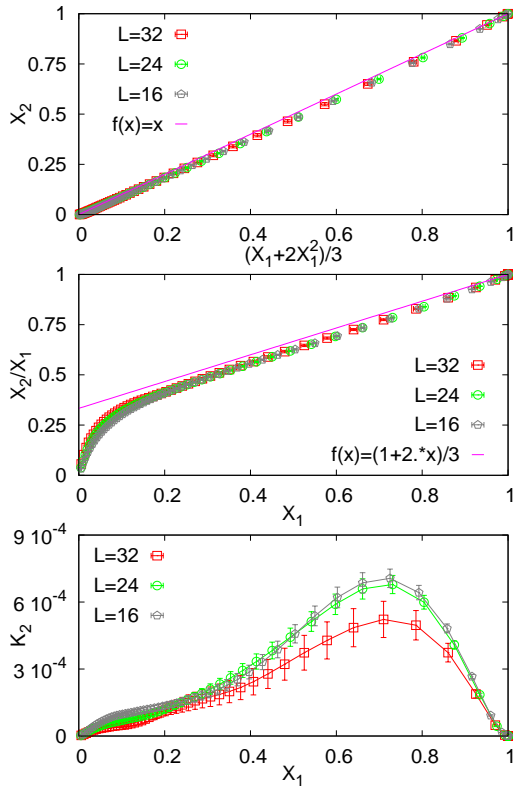}
\end{center}
\caption{(Color online) Top: $X_2$ as a function of the corresponding
  polynomial in $X_1$ (Eq.~(\ref{eq:X2SS})). The straight line is the
  theoretical prediction (unit slope).  Center: the ratio $X_2/X_1$ as
  a function of $X_1$, where the straight line is the theoretical
  prediction. Bottom: the squared difference
  $K_2=\left[X_2-(X_1+2X_1^2)/3\right]^2$ as function of $X_1$. Data
  refer to $T\sim 0.64T_\text{c}$}
\label{fig:X2}
\end{figure}

For models in which the overlap is not fluctuating in the large-volume
limit (i.e., $P(q)$ is a delta function) the above relations are
satisfied but reduce to trivial identities. If the replica symmetry is
broken, then stochastic stability imposes strong constraints on the
form of the overlap matrix and consequently on the overlap probability
densities. Ultrametricity is a further simplification: lack of this
property might indicate that more than one overlap might be needed to
describe the equilibrium configurations~\cite{PRJPA}.

We can extract further information from the distribution
$\Pi_{q}(x)$. It has been found~\cite{MPSTV,MPV,MPR} that in mean-field
theory the probability distribution $\pi(y)$ of the random
variable $Y_J=1-X_J$ behaves as a power law for $Y_J\sim 1$. This
implies that $\Pi_{q}(x)$ also follows a power law for small $x$
values
\bea
\label{eq:pixpow}
\pi_q(y \rightarrow 1) &\sim& (1-y)^{x(q)-1}\ \ , \nonumber \\
\Pi_q(s \rightarrow 0) &\sim& s^{x(q)-1}\ \ .
\eea
Since for most samples the $P_J(q)$ is a superposition of narrow peaks
around sample-dependent $q$ values, separated by wide $q$ intervals in
which $P_J$ is exactly zero, when dealing with data from simulations
of finite-size systems, it is convenient to turn to the cumulative
distribution of the $X_J$ to improve the statistical signal,
especially at small $q$ values:
\be
\label{eq:pixcum}
\Pi_q^C(s)=\int_0^sdx \Pi_q(x)
\ee
which should verify at small $s$
\be
\label{eq:pixcumpow}
\Pi_q^C(s \rightarrow 0) \sim s^{x(q)}\ ;
\ee
the probability of finding a sample in which the overlap probability
distribution $P_J(q)$ in the interval $\left[0,q\right]$ is small
enough to verify $\int_{-q}^qP(q^\prime)dq^\prime < s$ goes to zero as a
power law of $s$.

\subsection{Numerical results}
\label{subsec:numerics}

We recall that in our simulations we tailored the temperature range for the
parallel tempering implementation to improve its performance as discussed in
Ref.~\onlinecite{EAPTJANUS}. This brought us to direct measurements of
observables at temperature sets that were not perfectly overlapping at all
lattice sizes. In what follows we compare data at temperatures that are
slightly different for different lattice sizes. Considering that even if the
simulations were performed at exactly the same temperatures, tiny
size-dependent critical effects may always affect the results, we preferred
not to perform involved interpolations to correct for order $1\%$ or less of
temperature discrepancies. In what follows we will refer to the set of data at
$T\sim 0.64T_\text{c}$ and $T\sim 0.75T_\text{c}$ for the sake of brevity; the
precise values of the temperatures are summarized in Table~\ref{tab:Ts}. We
also compare data at exactly $T=0.625=0.57T_\text{c}$ for lattice sizes
$L=8,16,24$.

As our simulations were not optimized to study the critical region,
we take the value $T_\text{c}=1.109(10)$ from Refs.~\onlinecite{HPV1}
and~\onlinecite{HPV2} (featuring many more samples and small sizes to control
scaling corrections). Still, combining the critical exponents determination of
these references with the Janus data used herein, we obtain a
compatible value of $1.105(8)$.~\cite{DTC2}

\begin{table}[tb]
\begin{center}
\begin{tabular}{|c|c|c|c|}
\hline
$L$ & $T\sim 0.57T_\text{c}$ & $T\sim 0.64T_\text{c}$ & $T\sim 0.75T_\text{c}$ \\
\hline
8   &  0.625    &  $-$     &  0.815 \\
16  &  0.625    &  0.698    &  0.844 \\
24  &  0.625    &  0.697    &  0.842 \\
32  &  $-$      &  0.703    &  0.831 \\
\hline
\end{tabular}
\end{center}
\caption{Temperature values for each lattice size
  ($T_\text{c}=1.109$~\cite{HPV1,HPV2}).}
\label{tab:Ts}
\end{table}

\begin{figure}[tb]
\begin{center}
\includegraphics[width=0.8\columnwidth]{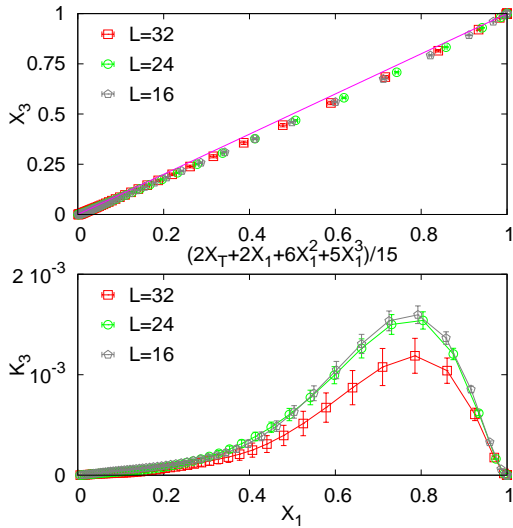}
\end{center}
\caption{(Color online) Data at $T\sim0.64T_\text{c}$. Top: $X_3$ as a
  function of the corresponding polynomial in $X_1$ and $X_\text{T}$
  (Eq.~(\ref{eq:X3SS})).  The straight line is the theoretical
  prediction (unit slope). Bottom: the squared difference
  $K_3=\left[X_3-(2X_\text{T}+2X_1+6X_1^2+5X_1^3 )/15\right]^2$ as function
  of $X_1$, $T=0.64T_\text{c}$. Lines connecting points are only a guide to
  the eye.}
\label{fig:X3}
\end{figure}

\begin{figure}[tb]
\begin{center}
\includegraphics[width=0.8\columnwidth]{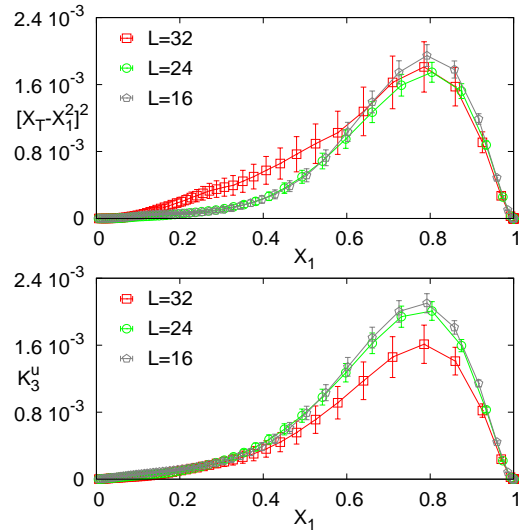}
\end{center}
\caption{(Color online) Top: The squared difference $\left[X_\text{T}-X_1^2
  \right]^2$ as a function of $X_1$. Bottom: the quantity
  $K_3^u=\left[X_3-(2X_1+8X_1^2+5X_1^3)/15\right]^2$ as a function of
  $X_1$. All data for $T\sim 0.64T_\text{c}$ and for lattice sizes
  $L=16,24,32$. The lines connecting the data points are only intended
  as a guide to the eye.}
\label{fig:ultra}
\end{figure}

We simulated four independent
real replicas per sample: thus we avoid any bias in computing $X_\text{T}(q)$,
Eq.~(\ref{eq:XT}), by picking three configurations in three distinct
replicas. We show the computed $X_\text{T}(q)$ for the largest lattices
$L=24$ and $L=32$ in Fig.~\ref{fig:XT} i) considering only
configurations for different replicas (data labeled as \emph{ABC});
ii) picking two configurations out of three from the same replica
(labeled \emph{AAB}); iii) picking the three configurations in the
same replica (labeled \emph{AAA}) . To minimize the effect of bias due
to \emph{hard} samples, we picked up the same number of configurations
per sample, spaced in time by an amount proportional to the
exponential autocorrelation time $\tau_\text{exp}$ of that
sample~\cite{EAPTJANUS}. The three data sets (ABC, AAB, AAA) are
equivalent and small deviations at low $q$ values remain within
error bars: this is a strong indication of the statistical quality of
our data, as described in Ref.~\onlinecite{EAPTJANUS}.

We now come to test the Ghirlanda-Guerra relations,
Eqs.~(\ref{eq:X2SS}) and (\ref{eq:X3SS}). Plotting the two sides of
Eq.~(\ref{eq:X2SS}) parametrically in $q$, the data show a slight
deviation from the theoretical prediction (see Fig.~\ref{fig:X2}
top). It is interesting to compare the discrepancies for different
lattice sizes. As the position and width of $P(q)$ are size-dependent,
it seems more natural to compare functions of the moments $X_k$ for
different lattice sizes as functions of the integrated probability
$x(q)=X_1(q)$ (see Fig.~\ref{fig:X2} middle). It is evident from the third plot in Fig.~\ref{fig:X2}
that the quantity
\be
K_2=\left[X_2-(X_1+2X_1^2)/3\right]^2
\ee
is definitely non-zero although very small in the entire range.
However, the data are compatible with $K_2$ decreasing with lattice
size and becoming null in the $L\to\infty$ limit.

We can reach similar conclusions regarding $X_3$ as a function of
$X_\text{T}$ and $X_1$, and the quantity
\be
\label{eq:K3}
K_3=\left[X_3-(2X_\text{T}+2X_1+6X_1^2+5X_1^3)/15\right]^2
\ee
(see Fig.~\ref{fig:X3}). Even if the data for different lattice sizes
stand within a couple of standard deviations, there is a clear
improvement in the agreement between the prediction and the Monte
Carlo data as the size increases.

The data plotted in Fig.~\ref{fig:ultra} take into account the
ultrametric relation~(\ref{eq:XTultra}). When comparing $X_\text{T}$ and
$X_1^2$ small deviations from the prediction arise. However, data for
$L=32$ have strong fluctuations, and do not hint at any
clear tendency with the system size.  The bottom plot in
Fig.~\ref{fig:ultra} shows data for the quantity
\be
\label{eq:K3u}
K_3^u=\left[X_3-(2X_1+8X_1^2+5X_1^3)/15\right]^2,
\ee
which we obtain by substituting (\ref{eq:XTultra}) in ~(\ref{eq:K3}).
The same considerations we made above apply here: the agreement with
ultrametric relations (\ref{eq:X3SS}) and (\ref{eq:XTultra}) improves
with increasing $L$.

\begin{figure}[tb]
\begin{center}
\includegraphics[width=0.97\columnwidth]{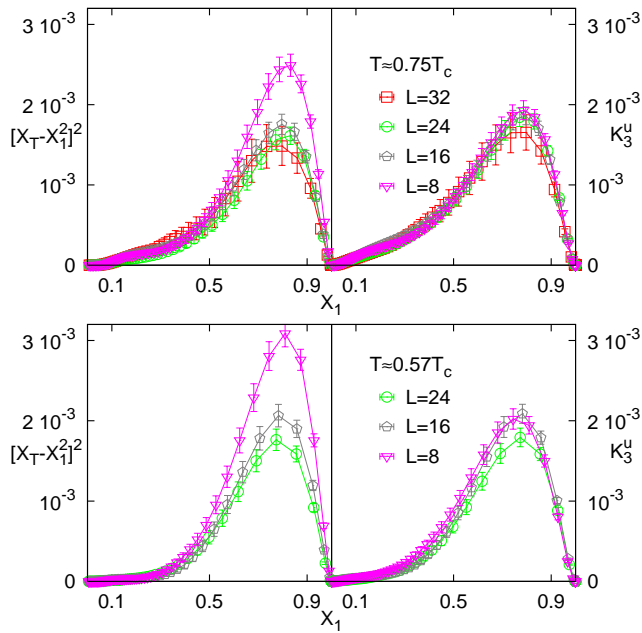}
\end{center}
\caption{(Color online) Square difference
  $\left[X_\text{T}-X_1^2 \right]^2$ (left) and the quantity
  $K_3^u=\left[X_3-(2X_1+8X_1^2+5X_1^3)/15\right]^2$ (right) as a
  function of $X_1$. Top: for $T=0.75T_\text{c}$ and $L=8, 16, 24,
  32$. Bottom: for $T=0.57T_\text{c}$ and $L=8, 16, 24$.}
\label{fig:0.75_0.57}
\end{figure}

\begin{figure}[tb]
\begin{center}
\includegraphics[width=0.8\columnwidth]{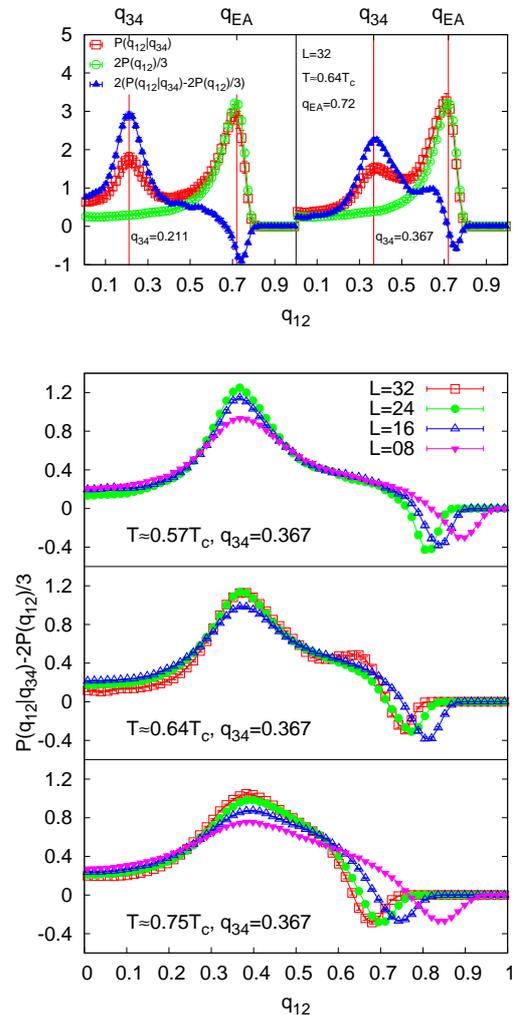}
\end{center}
\caption{(Color online) Top: The
  conditioned probability $P(q_{12}|q_{34})$ (open squares) for $L=32$
  and $T\sim0.64T_\text{c}$ and two values of $q_{34}=0.211$ (left) and
  $q_{34}=0.367$ (right). We also plot
  $2P(q_{12})/3$ (open circles) and the difference (full
  triangles) of the two above quantities (Eq.~(\ref{eq:pqqdiff}) in
  the text), scaled by a factor $2$ for a better view. $q_{34}$ and
  $q_\text{EA}$ values are indicated by vertical lines for visual
  reference. We took the value $q_\text{EA}(L=32,T=0.64T_\text{c})\sim0.72$ as
  given in Ref.~\onlinecite{EAPTJANUS}.\\ Bottom: The difference
  $P(q_{12}|q_{34})-2P(q_{12})/3$ with $q_{34}=0.367$, for different
  lattice size compared at temperatures $T=0.75T_\text{c}$, $T=0.64T_\text{c}$,
  $T=0.57T_\text{c}$.}
\label{fig:Pqq_PqPq}
\end{figure}

\begin{figure}[tb]
\begin{center}
\includegraphics[width=0.8\columnwidth]{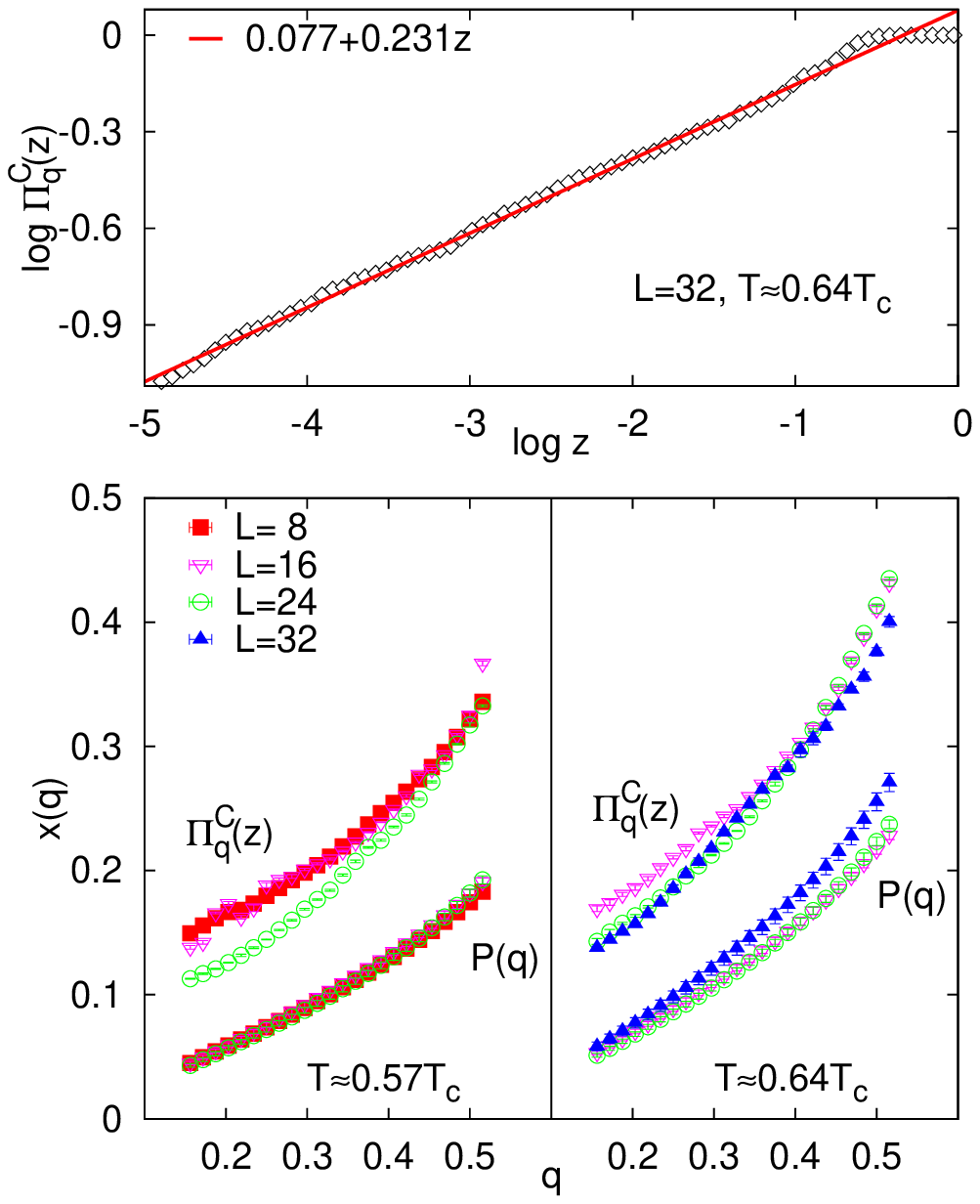}
\end{center}
\caption{(Color online) Asymptotic behavior of the cumulative
  probability $\Pi_{q}^C(z)$ (Eq.~(\ref{eq:pixcumpow})). Top:
  small-$z$ decay for $L=32$, $T=0.64T_\text{c}$ and $q=0.3125$. Bottom:
  comparison of the exponent $x(q)$ obtained by the two methods
  described in the text (uppermost data points represent values obtained
  by fitting $\Pi_{q}^C(z)$,  lowermost data points come from
  integrating the $P(q)$), for some lattice sizes, many cut-off values
  $q$ and temperatures $T\sim 0.57T_\text{c}$ (left) and $T\sim 0.64T_\text{c}$
  (right).}
\label{fig:Xqcompare}
\end{figure}

We can compare the results above with those of Ref.~\onlinecite{MPR},
in which a good agreement between theoretical prediction of the kind
of Eqs.~(\ref{eq:X2SS}), (\ref{eq:X3SS}), (\ref{eq:XTultra}) and Monte
Carlo data on $3D$ Edwards-Anderson spin glass with Gaussian couplings
was reported, but without clear evidence on whether the very small
discrepancies could be controlled or not in the limit of large
volume. In this respect, we have been able to thermalize systems of
linear sizes up to twice the largest lattice studied in
Ref.~\onlinecite{MPR} and these larger sizes show a trend towards
satisfying Eqs.~(\ref{eq:X2SS}), (\ref{eq:X3SS}), (\ref{eq:XTultra})
that was not clear in Ref.~\onlinecite{MPR}. We also note that
finite-size effects are stronger at low temperatures, and obtaining
evidence of the correct trend requires data from simulations of larger systems
than at higher temperature. We can also compare data at $T\sim 0.75T_\text{c}$
and $T=0.57T_\text{c}$ (we have data at exactly $T=0.625$ for lattice sizes
$L=8$, $L=16$ and $L=24$ but unfortunately not for $L=32$). We see
that at $T\sim 0.75T_\text{c}$ the data for the squared differences $K_3^u$ and
$\left[X_\text{T}-X_1^2 \right]^2$ are almost size-independent (this is
actually true for $\left[X_\text{T}-X_1^2 \right]^2$ when $L>8$, see
Fig.~\ref{fig:0.75_0.57}, top). At $T\sim 0.64T_\text{c}$ (see
Fig.~\ref{fig:ultra}), such effects cannot be clearly told by comparing
only the smallest lattices considered, $L=16$ and $L=24$. At
$T=0.57T_\text{c}$, size-dependent effects are strong even for $L=16,24$ (see
Fig.~\ref{fig:0.75_0.57}, bottom).

Having data from four independent replicas per sample, we have access
to the joint probability of two independent overlaps.  According to
Eq.~(\ref{eq:PqSS1}) the quantity
\be
\frac{P(q_{12},q_{34})}{P(q_{34})} - \frac{2}{3} P(q_{12}) =
P(q_{12}|q_{34})-\frac{2}{3} P(q_{12})\ ,
\label{eq:pqqdiff}
\ee
(where $P(\cdot|\cdot)$ denotes conditional probability)
when plotted versus $q_{12}$, should be a delta function in $q_{34}$.
This quantity is shown for $L=32$, $T\sim0.64T_\text{c}$ and two values of
$q_{34}$ in the top plot of Fig.~\ref{fig:Pqq_PqPq} and reveals a
clear peak around $q_{34}$. At high $q_{12}$ values there is a small
excess in the probability $P(q_{12})\,P(q_{34})$, so the
difference in Eq.~(\ref{eq:pqqdiff}) becomes negative. As one sees in
Fig.~\ref{fig:Pqq_PqPq} this happens at values $q_{12} \gtrsim
q_\text{EA}$, i.e., in a region of atypically large overlaps that should
vanish in the thermodynamical limit. The size dependence for the quantity
in Eq.~(\ref{eq:pqqdiff}) is not easy to quantify from the data: as
one can see in Fig.~\ref{fig:Pqq_PqPq} (bottom) for a particular
choice of $q_{34}$, the peak height tends to increase with $L$ (at
least for $T \sim 0.75T_\text{c}$), but in a very slow way, making
extrapolations in the $L\to\infty$ limit practically impossible.
Despite this, we note that the negative peaks get narrower as
the system size increases: we expect then that this effect will disappear
at larger system sizes.

We conclude this section commenting the asymptotic behavior of the
cumulative probability $\Pi_{q}^C(z)$, Eq.~(\ref{eq:pixcumpow}). The
small-$z$ decay is clearly a power law (see top plot in
Fig.~\ref{fig:Xqcompare}), but the best fit exponent is significantly
different from the estimate obtained by integrating the overlap
distribution $P(q)$. Fig.~\ref{fig:Xqcompare} shows a comparison of
the exponent $x(q)$ obtained by the two methods, for some lattice
sizes, many cut-off values $q$ and two temperatures, $T\sim 0.64T_\text{c}$
and $T\sim 0.57T_\text{c}$.  Although the differences seem to decrease by
increasing the lattice size, the trend is very slow and even not in a
clear direction for some values of the cutoff $q$.  Again, the only
conclusion that can be drawn is that the finite-size effects are large,
even for $L=32$, and safe extrapolations in the $L\to\infty$ limit cannot
be done.

A closer inspection of the data reported in Fig.~\ref{fig:Xqcompare}
reveals that the difference between the two data sets is roughly a
constant, and this difference becomes extremely important in the limit
of small $q$, where one would expect both measurements of $x(q)$ to
approach zero. Contrary to expectations, the $x(q)$ estimated from
the data of $\Pi^C_{q}$ seems to remain non-zero even in the $q\to 0$
limit.  A possible explanation for this observation comes from the
fact that the delta peaks in the $P_J(q)$ get broader for systems of
finite size.  Indeed, in the thermodynamic limit, one would expect
$P_J(q)$ to be the sum of delta functions centered on overlap values
extracted from the average distribution $P_\infty(q)$: if this
expectation is true, then the value for $X_J(q)$ is nothing but the
probability of having a peak at an overlap value smaller than $q$ and
this is exactly $x(q)$.  However, if the delta peaks acquire a non-zero
width $\Delta$ due to finite-size effects, then for $q<\Delta$
the overlap probability distribution close to the origin $P_J(0)$ may
be affected by broad peaks centered on overlaps larger than $q$,
which should not count in the thermodynamical limit.  If this
explanation is correct, then the limit $q \to 0$ for the data shown
in Fig.~\ref{fig:Xqcompare} (bottom) obtained from $\Pi^C_{q}$
should give a rough estimate, in the large $L$ limit, for the peak
width $\Delta$ (see data in Table~\ref{tab:fitk2.5} and discussion
below).

\section{THE ORDER PARAMETER DISTRIBUTION}
\label{sec:pq}

\begin{table}[tb]
\begin{center}
\begin{tabular}{|c|c|c|c|c|}
\hline
$L$ & $T/T_\text{c}$ & $q_\text{EA}$ & $x_\infty(q_\text{EA})$  & $\Delta$ \\
\hline
32 & 0.75 & 0.663(19)   & 0.91(13)   & 0.0923(80)  \\   
   & 0.64 & 0.7319(30)  & 0.828(28)  & 0.1015(30)  \\ 
24 & 0.75 & 0.69674(72) & 1.0000(3)  & 0.10618(84)  \\ 
   & 0.64 & 0.7625(27)  & 0.876(24)  & 0.1182(24)  \\ 
   & 0.57 & 0.7954(24)  & 0.842(25)  & 0.1216(32)  \\ 
16 & 0.75 & 0.73780(73) & 1.000031(7)& 0.1443(10)  \\ 
   & 0.64 & 0.809(16)   & 1.00(14)   & 0.150(11)  \\
   & 0.57 & 0.8210(41)  & 0.811(49)  & 0.1683(51)   \\
8  & 0.75 & 0.8250(21)  & 1.000001(9)& 0.2872(37)   \\
   & 0.57 & 0.886(18)   & 0.95(18)   & 0.296(28)  \\
\hline
$L$ & $T/T_\text{c}$ & $\alpha$ & $\gamma$ & $\chi^2/\text{d.o.f.}$\\
\hline
32 & 0.75 & 1.92(34) & 11.2(1.2) & 20/97 \\ 
   & 0.64 & 0.93(44) & 7.7(1.0)  & 38/103 \\ 
24 & 0.75 & 2.04(21) & 9.68(55)  & 45/101 \\ 
   & 0.64 & 0.95(21) & 6.88(41)  & 69/107 \\ 
   & 0.57 & 0.75(17) & 5.62(30)  & 88/110 \\ 
16 & 0.75 & 1.76(16) & 5.14(31)  & 77/107 \\ 
   & 0.64 & 0.45(21) & 4.50(52)  & 133/113 \\
   & 0.57 & 0.53(19) & 3.37(40)  & 161/115 \\
8  & 0.75 & 0.73(22) & 2.02(34)  & 501/121 \\
   & 0.57 & 0.49(16) & 1.36(17)  & 466/123 \\
\hline
\end{tabular}
\end{center}
\caption{Results of the fitting procedure of Eq.~(\ref{eq:conv}) on
  numerical $P(q)$ data, with kernel exponent $k=2.5$ (see
  Eq.~(\ref{eq:G(h,k)})). All errors on parameters are jackknife
  estimates.~\footnote{We used the symbol $\chi^2$ in the table to denote the
    sum of squares of residuals, which is not a true chi-square estimator as
    the values of $P(q)$ at different $q$ are mutually correlated.}}
\label{tab:fitk2.5}
\end{table}


\begin{figure}
\begin{center}
\includegraphics[width=0.8\columnwidth]{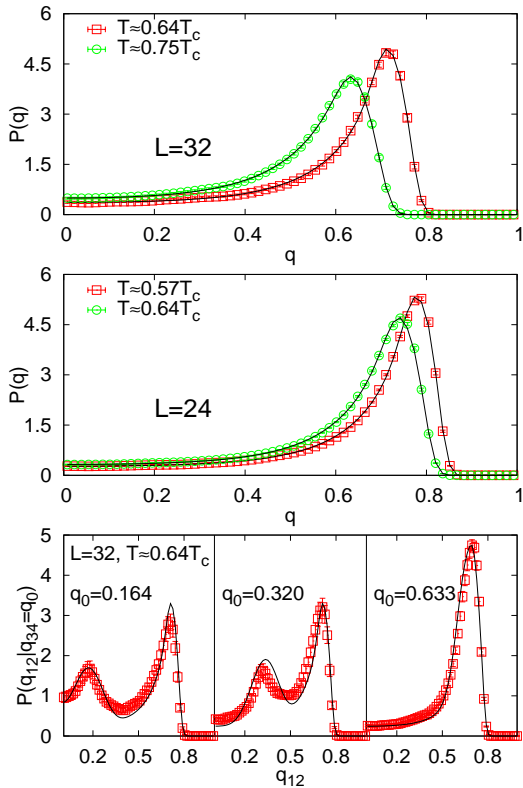}
\end{center}
\caption{(Color online) Comparison between the Monte Carlo data of the
$P(q)$ and the convolution computed as described in the text (solid
  lines). Top: $L=32$, $T\sim0.64T_\text{c}$ and $T\sim 0.75T_\text{c}$. Center:
  $L=24$, $T\sim0.57T_\text{c}$ and $T\sim0.64T_\text{c}$. Bottom: the conditioned
  probability $P(q_{12}|q_{34}=q_0)$ for $L=32$, $T\sim 0.64T_\text{c}$ and
  some values of $q_0$.  }
\label{fig:fitk2.5}
\end{figure}

We now compare the $P(q)$ obtained in numerical simulations of the
three-dimensional Edwards-Anderson model~(\ref{eq:EA}) to the prediction
obtained by smoothly introducing controlled finite-size effects on a
mean-field-like distribution consisting in a delta function centered in
$q=q_\text{EA}$ and a continuous tail down to $q=0$ (a similar analysis has been carried out for long-range spin-glass models, see Ref.~\onlinecite{LR}). On the positive $q$ axis one
has
\bea
P_\infty (q) & = & \widetilde{P}(q)\Theta(q_\text{EA}-q) \nonumber \\
            & + & [1-x_\infty(q_\text{EA})]\delta(q-q_\text{EA})\;,\\
x_\infty(q_\text{EA}) & = & \int_0^{q_\text{EA}}dq\,\widetilde{P}(q)\;.
\label{eq:PqMF}
\eea
It is convenient to introduce the effective field $h$ trough
\be
\label{eq:qtanh}
q=\tanh \left(h\right)
\ee
and consider its distribution
\bea
\mathcal{P}_\infty(h) & = & P_\infty \big(q(h)\big) \frac{dq(h)}{dh} \nonumber\\
&=& \frac{dq(h)}{dh} \widetilde{P}\big(q(h)\big) \Theta(h_\text{EA}-h) + \nonumber \\
&& [1-x_\infty(q_\text{EA})] \delta(h-h_\text{EA})\;, \label{eq:Ph} \\
x_\infty(q_\text{EA})&=&\int_0^{h_\text{EA}}dh\widetilde{\mathcal{P}}(h)\;,
\eea
being clear that $q_\text{EA}=\tanh \left(h_\text{EA}\right)$.
This change of variable smooths the constraint on the fluctuations of $q$
near the extremes of the distribution.

In a finite-size system the thermodynamical distribution
$\mathcal{P}_\infty(h)$ will be modified, mainly by the fact that
delta functions become distributions with non-zero widths.  Remember
that, in the thermodynamical limit, we expect the distribution
$\mathcal{P}_J(h)$ for any given sample to be the sum of delta
functions.  A simple way to take into account the spreading of the
delta functions due to finite-size effects is to introduce a symmetric
convolution kernel
\be
\label{eq:G(h,k)}
G^{(k)}_\Delta(h-h^\prime) \equiv C\exp{\left[ -\left(\left| h-h^\prime
    \right|/\Delta\right)^k\right]}\;,
\ee
where $C$ is a normalizing constant and the spreading parameter
$\Delta$ is assumed not to depend on $h$,~\footnote{This introduces a
  $q$-dependent spread, as the Jacobian of the
  transformation~(\ref{eq:qtanh}) stretches the distribution at high
  $q$ values.} while it should have a clear dependence on the system
size, such that $\lim_{L\to\infty}\Delta=0$.  The parameter $k$, to be
varied in the interval $[2,3]$, is introduced in order to consider
convolutions different from the Gaussian case ($k=2$).

In order to obtain an analytic expression for the finite
size distribution
\be
\mathcal{P}_L(h) \equiv \int dh' \frac{\mathcal{P}_\infty(h') +
\mathcal{P}_\infty(-h')}{2} G^{(k)}_\Delta(h-h')\;,
\ee
we assume the following form for the continuous part of the
distribution
\be
\widetilde{\mathcal{P}}(h) \equiv \widetilde{P}\big(q(h)\big)
\frac{dq(h)}{dh} = \widetilde{P}(0)(1+\alpha h^2 + \gamma h^4)\;,
\label{eq:Phpol}
\ee
where $ \widetilde{\mathcal{P}}(0) = \widetilde{P}(0) = P_\infty(0)$,
$\alpha$ and $\gamma$ are free parameters to be inferred from the
data.  The final result is
\bea
\mathcal{P}_L(h) & = & [1-x_\infty(q_\text{EA})]
\frac{G^{(k)}_\Delta(h-h_\text{EA}) + G^{(k)}_\Delta(h+h_\text{EA})}{2}
\nonumber \\
& + & \widetilde{P}(0)\int_{-h_\text{EA}}^{h_\text{EA}}\!\!\!\!\!\!dz\,
\left[1+\alpha z^2 + \gamma z^4 \right]
G^{(k)}_\Delta(h-z) \label{eq:conv}
\eea
where $x_\infty(q_\text{EA})=2 \widetilde{P}(0) [h_\text{EA} +
\alpha h_\text{EA}^2/3 + \gamma h_\text{EA}^5/5]$.

We let $\alpha$, $\gamma$, $q_\text{EA}$ and $\Delta$ vary in a fitting
procedure to $P(q)$ Monte Carlo data; values of $\widetilde{P}(0)$ are
fixed to the Monte Carlo values $P_{MC}(0)$. The choice of the
exponent $k$ in the convolution kernel is crucial. We varied $k$ in
the interval $\left[2,3\right]$. The Gaussian convolution $k=2$ turned
out to be the worst choice in such interval, giving rise to unphysical
negative weights for the delta function contributions,
i.e., $1-x_\infty(q_\text{EA})<0$. We obtained very good results with the
choice $k=2.5$. Fit parameters are reported in
Table~\ref{tab:fitk2.5} for some lattice sizes and temperatures, while
Fig.~\ref{fig:fitk2.5} shows comparison between Monte Carlo $P(q)$ and
the relative fitting curve.  Although the fitting curves interpolate
nicely the numerical $P(q)$, some of the fitting parameters may look
strange: in particular $q_\text{EA}$ is a bit larger than the peak
location and $x_\infty(q_\text{EA}) \simeq 1$ (for example, in the $L=32$ data
the difference is around $2\%$).  It is worth remembering
that in the solution of the SK model at low temperatures the continuous
part $\mathcal{P}(q)$ has a divergence for $q \to q_\text{EA}^-$, which can
easily dominate the delta function in finite-size systems (where delta
peaks are broadened).  Indeed, by increasing the system size, $q_\text{EA}$
seems to move towards the location of the peak maximum and
$x_\infty(q_\text{EA})$ becomes smaller than 1.

In order to make a stronger test of the above fitting procedure, we
have used the fit parameters in Table~\ref{tab:fitk2.5} to
derive the finite-size conditional probability
\be
P_{L}(q|q^\prime) = P_L(q,q^\prime)/P_L(q^\prime)
\label{eq:PqinfSS1}
\ee
applying the convolution kernel
$G^{(k)}_\Delta(h-h')$ to the $L=\infty$ joint probability given by
the Ghirlanda-Guerra relation, r.h.s of Eq.(\ref{eq:PqSS1}).
Fig.~\ref{fig:fitk2.5} shows a comparison between our extrapolated
$P_L(q_{12}|q_{34}=q_0)$ and the Monte Carlo data for $L=32$,
$T=0.64T_\text{c}$ and three values of $q_0$: the agreement is very good at
any value of $q_0$, especially considering that the fitting parameters
were previously fixed by interpolating the unconditional overlap
distribution $P_L(q)$.

\section{CONCLUSIONS}
\label{sec:conclusions}

We performed a direct inspection of stochastic stability and
ultrametricity properties on the sample-to-sample fluctuations of the
overlap probability densities obtained by large-scale Monte Carlo
simulations of the three-dimensional Edwards-Anderson model. We found
small but still sizeable deviations from the prediction of the
Ghirlanda-Guerra relations but a clear tendency towards improvement of
agreement with increasing system size.

Large fluctuations make it difficult to draw any definitive conclusion
on the analysis of the ultrametric relation~(\ref{eq:XTultra}) when
taking into account data for the largest lattice size. In addition,
critical effects show up at $T\sim 0.75T_\text{c}$.  Considering that for a
stochastically stable system overlap equivalence is enough to
infer ultrametricity, the results presented here support and integrate
the analyses and claims of Refs~\onlinecite{janusPRL},
~\onlinecite{EAPTJANUS} and ~\onlinecite{ContucciPRL}, in which the
authors reported strong evidence of overlap equivalence.

We also turned our attention to the shape of the overlap probability
distribution, showing that finite-size $P_L(q)$ and $P_L(q,q')$
compare well with mean-field (infinite-size) predictions, modified by
finite-size effects that only make delta functions broad.

\acknowledgments
Janus has been funded by European Union (FEDER) funds, 
Diputaci\'on General de Arag\'on (Spain), 
by a Microsoft Award-Sapienza-Italy, and by Eurotech.
We acknowledge partial financial support from MICINN,
Spain, (contracts FIS2009-12648-C03, FIS2010-16587, TEC2010-19207), Junta de Extremadura
(GR10158), UEx (ACCVII-08) and from UCM-Banco de Santander 
(GR32/10-A/910383). D. I\~niguez is supported by the Government
of Aragon through a Fundaci\'on ARAID contract. 
B. Seoane and D. Yllanes are supported by the FPU program
(Ministerio de Educaci\'on, Spain).



\end{document}